\documentclass[acmsmall]{acmart}

\AtBeginDocument{%
  \providecommand\BibTeX{{%
    \normalfont B\kern-0.5em{\scshape i\kern-0.25em b}\kern-0.8em\TeX}}}

\setcopyright{acmcopyright}
\copyrightyear{2021}
\acmYear{2021}

\begin{document}

\title{Mixed Reality Interaction Techniques}

\author{Jens Grubert}
\email{jg@jensgrubert.de}

\renewcommand{\shortauthors}{Grubert}

\begin{abstract}
This chapter gives an overview of interaction techniques for mixed reality including augmented and virtual reality (AR/VR). Various modalities for input and output are discussed. Specifically, techniques for tangible and surface-based interaction, gesture-based, pen-based, gaze-based, keyboard and mouse-based, as well as haptic interaction are discussed. Furthermore, the combination of multiple modalities in multisensory and multimodal interaction as well as interaction using multiple physical or virtual displays are presented. Finally, interaction with intelligent virtual agents is considered.
\end{abstract}

\maketitle


\section{Introduction}



This chapter gives an overview of interaction techniques for mixed reality (MR), including augmented and also virtual reality (AR/VR). Early research in the field of MR interaction techniques focused on the use of surface-based, tangible as well as gesture-based interaction, which will be presented at the beginning of this chapter. Further modalities, such as pen-based, gaze-based or haptic interaction have seen recent attention and are presented next. Further, with the move towards productivity-oriented use cases, interaction with established input devices such as keyboard and mice has seen interest in the research community. Finally, inspired by the popularity of conversational agents, interaction with intelligent virtual agents are discussed. The development of interaction techniques is closely related to the advancements in input devices. Hence, the reader is invited to study the according book chapter as well. While this chapter follows the above mentioned structure, further possibilities to structure interaction techniques include organizing according to interaction tasks \cite{bowman20043d} such as object selection \cite{dang2007survey, argelaguet2013survey, singh2019object} and object manipulation \cite{mendes2019survey}, navigation \cite{jankowski2013survey}, symbolic input \cite{dube2019text} or system control \cite{dachselt2007three}. Further, interaction techniques for specific application domains have been discussed such as music \cite{berthaut20193d} games \cite{riecke20183d} or immersive analytics \cite{buschel2018interaction, ens2021grand}. Interested readers are also referred to further surveys on 3D interaction techniques \cite{hand1997survey} or interaction with smart glasses \cite{lee2018interaction}.


\section{Tangible and Surface-based Interaction}

\begin{figure}
\includegraphics[width=0.8\columnwidth]{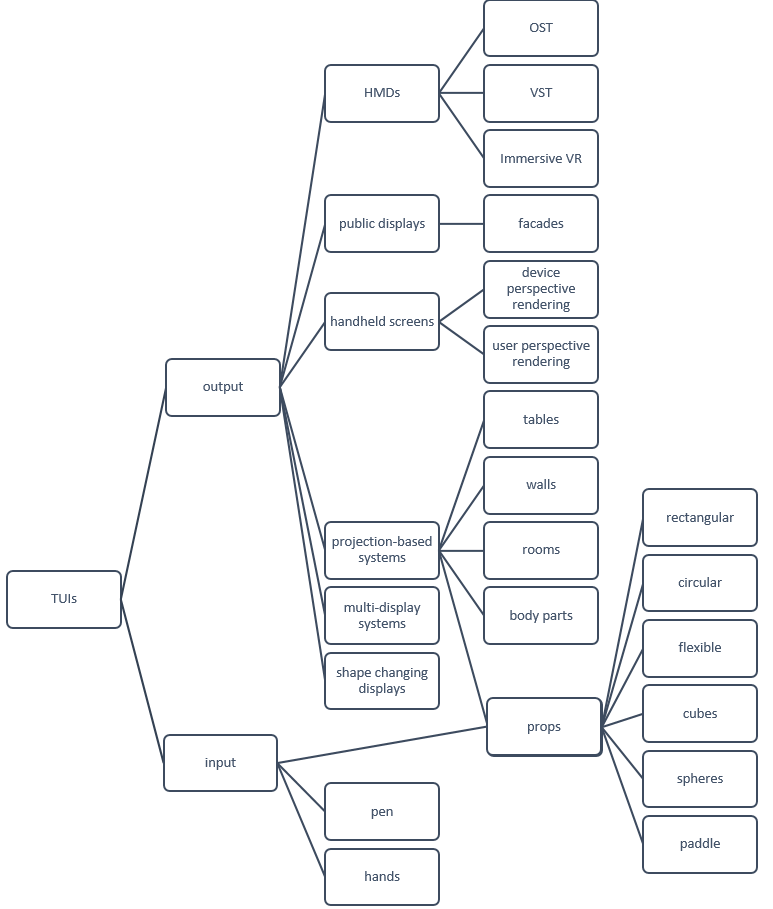}
\caption{Classification of input and output devices used in tangible user interfaces for AR and VR. OST: Optical See-Through. VST: Video See-Through. HMDs: Head-Mounted Displays.}
\label{fig:tuiclassification}
\end{figure}

This section presents the concepts of Tangible user interfaces (TUIs) and their applicability in AR. It covers the effects of output media, spatial registration approaches for TUIs, tangible magic lenses, augmenting large surfaces like walls and whole rooms, the combination of AR with shape-changing displays as well as the role of TUIs for VR-based interaction. Figure \ref{fig:tuiclassification} depicts an overview about output and input devices typically found in TUI-based interaction for MR.

TUIs are concerned with using physical objects as medium for interaction with computers \cite{ullmer1997metadesk} and have seen substantial interest in human-computer interaction \cite{shaer2010tangible}. Early prototypes utilized tabletops, on which physical objects were placed to change properties of digital media. For example Underkoffler and Ishii introduced a simulation of an optical workbench using tangible objects on a tabletop \cite{underkoffler1998illuminating} as well as an application for architectural planning \cite{underkoffler1999urp}.

In AR, this concept was introduced by Kato et al. \cite{kato2000virtual} as Tangible Augmented Reality (TAR). They used a paddle as prop, equipped with a fiducial, to place furniture inside a house model. Fjeld et al. \cite{fjeld2002augmented} introduced further tangibles such as a booklet and a cube for interacting within an educational application for chemistry.

TAR is typically used for visualizing digital information on physical objects while concurrently using those physical objects as interaction devices. Billinghurst et al. \cite{billinghurst2005designing} state TAR characteristics as realizing spatial registration between virtual and physical objects and the ability for users to interact with those virtual objects by manipulating the physical ones. For example, Regenbrecht et al. \cite{regenbrecht2002magicmeeting} utilized a rotary plate to allow multiple co-located users to manipulate the orientation of a shared virtual object. This way, the gap between digital output (e.g., on a flat screen) and physical input (e.g., using a rotary knob) can be reduced as the digital information is directly overlaid over the physical content. Lee et al. \cite{lee2007interaction} described common interaction themes in TAR applications such as static and dynamic mappings between physical and digital objects. They describe a space-multiplexed approach, where each physical tool is mapped to a single virtual tool or function as well as a time-multiplexed approach in which the physical object is mapped to different digital tools dependent on the context-of use.
However, the effect of this overlay depends also on the output medium. For example, when using projection-based systems \cite{bimber2005spatial} or video see-through (VST) head-mounted displays (HMDs) (c.f. chapter 10 in \cite{hainich2016displays}), the distance between the observer to the physical and virtual objects is the same. In contrast, when using commodity optical see-through (OST) HMDs with a fixed focal plane, there can be a substantial cost of perceiving virtual and physical objects at the same time. Specifically, Eiberger et al. \cite{eiberger2019effects} showed that when processing visual information jointly from objects within arms reach (in this case a handheld display) and information presented on a OST HMD at a different distance, task completion times increased by approximately 50\% and error rate increased by approximately 100\% compared to processing this visual information solely on the OST HMD.

For spatially registering physical and virtual objects, early works on TAR often relied on fiducial markers, such as provided by ARToolKit \cite{kato2007inside} or ARUCO \cite{salinas2012aruco}. While, easy to prototype (i.e. fiducials have to be simply printed out and attached to objects), these markers can inhibit interaction due to their susceptibility to occlusions (typically through hand and finger interaction). Hence, it is advised to use modern approaches for hand-based interaction \cite{barsoum2016articulated,li2019survey} with spatially tracked rigid and non-rigid objects \cite{rahman2019recent, guo2019deep, wu2019tangible}. Also, when using OST HMDs, the calibration between the HMD and the users' eyes can impact the interaction \cite{grubert2010comparative, grubert2017survey}.


Evolving from the magic lens \cite{bier1993toolglass} and tangible interaction concepts \cite{ullmer1997metadesk} \textit{tangible magic lenses} allow to access and manipulate otherwise hidden data in interactive spatial environments. A wide variety of interaction concepts for interactive magic lenses have been proposed within the scope of information visualization (c.f. \cite{tominski2014survey,tominski2017interactive}). 

Within AR, various rigid shapes have been explored. Examples include rectangular lenses for tabletop interaction \cite{spindler2009paperlens} or circular lenses \cite{spindler2010tangible}. Flexible shapes  \cite{steimle2013flexpad} have been utilized as well as multiple sheets of paper \cite{holman2005paper}. In their pioneering work, Szalavári and Gervautz \cite{szalavari1997personal} introduced the personal-interaction-panel in AR. The two-handed and pen-operated tablet allowed for selection and manipulation of virtual objects as well as for system control. Additionally, transparent props have been explored (e.g., a piece of plexiglass) both for tabletop AR \cite{rekimoto2001datatiles, brown2006magic, oh2006user} as well as VR \cite{schmalstieg1999using}. Purely virtual tangible lenses have been proposed as well \cite{looser20073d}. Brown et al. \cite{brown2003widget} introduced a cubic shape which could either perspectively correct render and manipulate 3D objects or text. This idea was later revisited by Issartel et al. \cite{issartel2016tangible} in a mobile setting. 

Often, projection-based AR has been used to realize tangible magic lenses, in which a ceiling-mounted projector illuminates a prop such as a piece of cardboard or other reflective materials \cite{spindler2009paperlens,chan2012magicpad} and (typically RGB or depth) cameras process user input.

\begin{figure}
\includegraphics[width=\columnwidth]{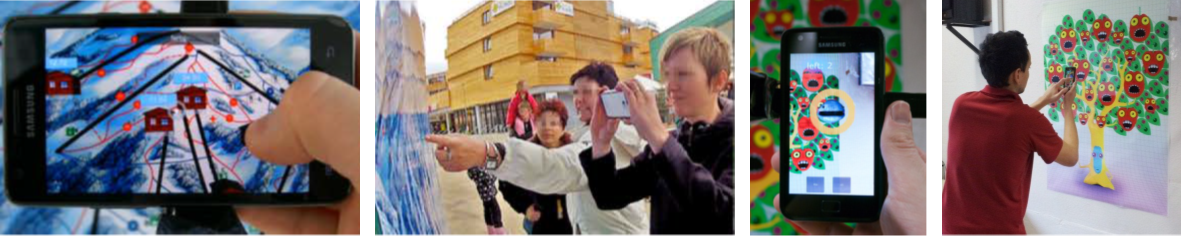}
\caption{Two magic lens applications in handheld AR. from left to right: Smartphone-based magic lens interface reveals virtual huts and prices on a physical ski map \cite{grubert2015utility}. Tourist searches for prices in front of the map. Find and select game using a handheld magic lens interface. User playing the game in front of a physical poster. \cite{grubert2012playing}}
\label{fig:magiclens}
\end{figure}

Mobile devices such as smartphones and tablets are also commonly used as a tangible magic lens \cite{grubert2015utility,leigh2015thaw}, see Figure \ref{fig:magiclens}, and can be used in conjunction with posters \cite{grubert2015utility}, books \cite{kljun2019augmentation}, digital screens \cite{leigh2015thaw} or maps \cite{reitmayr2005localisation, morrison2009like}.

When using the tangible magic lens metaphor in public space, one should be aware about the social acceptability, specifically due to the visibility of spatial gestures and postures \cite{reeves2005designing, rico2010usable}. For example, in a series of studies across gaming and touristic use cases, Grubert et al. \cite{grubert2012playing, grubert2013playing} explored benefits and drawbacks of smartphone-based tangible lens interfaces in public settings and compared them to traditional static peephole interaction, commonly used in mobile map applications. They found that user acceptance is largely dependent on the social and physical setting. In a public bus stop inside a large open space used at transit area, participants favored the magic lens over a static peephole interface despite tracking errors, fatigue and potentially conspicuous gestures. Also, most passersby did not pay attention to the participants and vice versa. However, when deploying the same experience in a different public transportation stop with other spatial and social context (waiting area, less space to avoid physical proximity to others), participants used and preferred the magic lens interface significantly less compared to a static peephole interface. In the context of of evaluations of magic lens metaphors in handheld AR, the impact of tracking technology also has to be considered \cite{mulloni2012experiences}.

\begin{figure}
\includegraphics[width=\columnwidth]{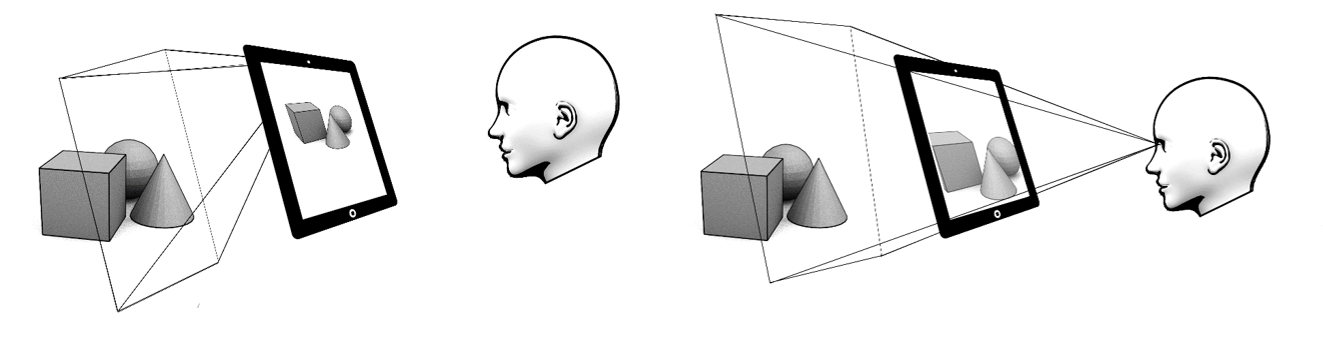}
\caption{In device-perspective rendering (left) the scene is rendered from the point of view of the camera. In user perspective rendering (right), the scene is rendered from the point of view of the user's head \cite{mohr2017adaptive}.}
\label{fig:dprupr}
\end{figure}

\begin{figure}
\includegraphics[width=0.5\columnwidth]{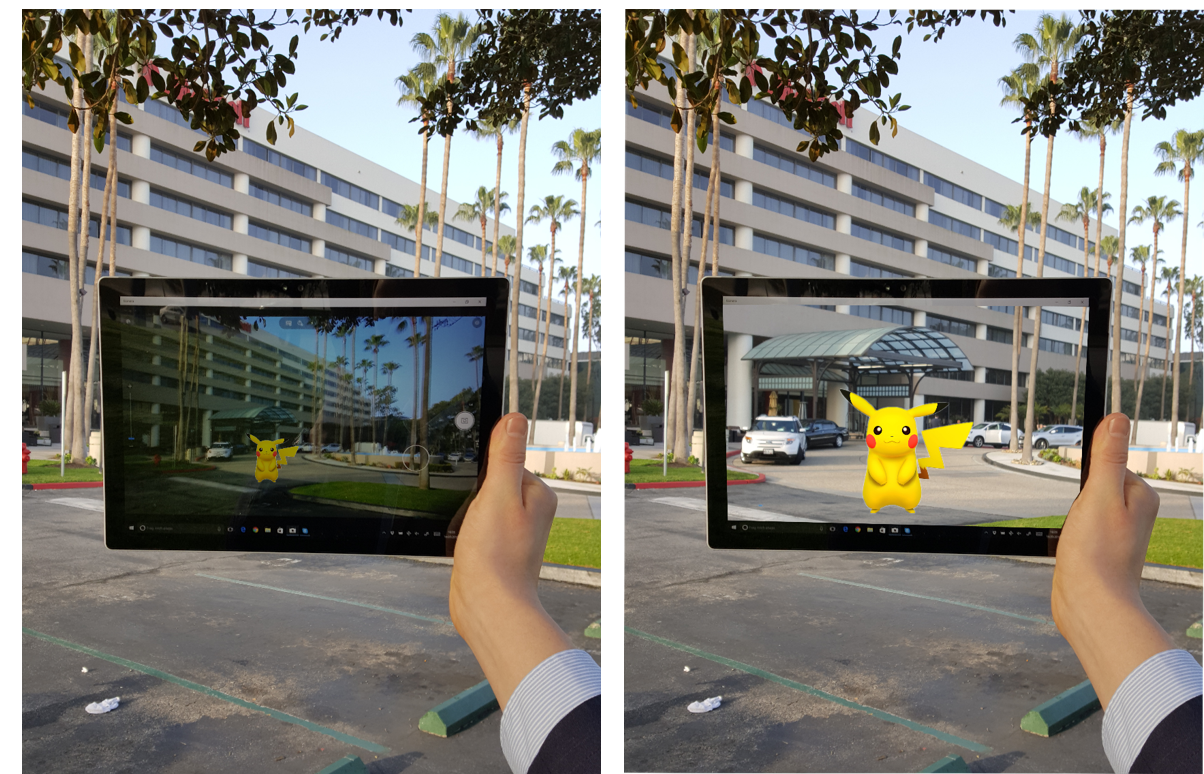}
\caption{Effects of device-perspective rendering (left) and user perspective rendering (right) \cite{mohr2017adaptive}.}
\label{fig:upreffects}
\end{figure}

Further, when using smartphones or tablets as magic lenses, the default user's view is based on the position of the physical camera attached to the handheld device, see Figure \ref{fig:dprupr}, left and Figure \ref{fig:upreffects}, left. However, this can potentially negatively affect the user experience \cite{vcopivc2013evaluating, vcopivc2014use}. Hence, it can be advisable to incorporate user-perspective rendering to render the scene from the point of view of the user's head, see Figure \ref{fig:dprupr}, right and Figure \ref{fig:upreffects}, right. For example, Hill et al. \cite{hill2011virtual}, introduced user-perspective rendering as \textit{virtual transparency} for VST AR. Bari{\v{c}}evi{\'c} et al \cite{barivcevic2012hand} compared user- vs. device-perspective rendering in a VR simulation. Tomioka et al. \cite{tomioka2013approximated} presented approximated user-perspective rendering using homographies. Grubert et al. \cite{grubert2014towards} proposed a framework for enabling user-perspective rendering for augmenting public displays. {\v{C}}opi{\v{c}} et al. \cite{vcopivc2013evaluating, vcopivc2014use}, quantified the performance differences between device- and user perspective rendering in map-related tasks and Mohr et al. \cite{mohr2017adaptive}, developed techniques for efficient computation of head-tracking techniques needed for user-perspective rendering.

Beyond handheld solutions, whole surfaces such as tables, walls or body parts can be augmented and interacted with. Often projector-camera systems are used for processing input and creating output on surfaces. Early works included augmenting desks using projectors to support office work of single users \cite{wellner1991digitaldesk, wellner1993interacting, mackay1999designing} or in collaborative settings \cite{rekimoto1999augmented}. Later, the Microsoft Kinect and further commodity depth sensors gave rise to a series of explorations with projector-camera systems.
For example, Xiao et al. \cite{xiao2013worldkit} introduced WorldKit, to allow users to sketch and operate user interface elements on everyday surfaces. Corsten et al. \cite{corsten2013instant} proposed a pipeline for repurposing everyday objects as input devices. Henderson and Feiner also proposed to utilize passive haptic feedback from everyday objects to interact with virtual control elements such as virtual buttons \cite{henderson2010opportunistic}. Mistry and Maes \cite{mistry2009sixthsense} utilized a necklace-mounted projector-camera system to sense finger interactions and project content on hands or the environment. Following suit, Harrison et al., \cite{harrison2011omnitouch} introduced OmniTouch, a wearable projector-depth-camera system that allowed to project user interface elements on body parts, such as the hand (e.g., a virtual dial pad) or to augment paper using touch. 

Beyond handheld solutions, whole surfaces such as tables, walls or body parts can be augmented and interacted with. Often projector-camera systems are used for processing input and creating output on surfaces. Early works included augmenting desks using projectors to support office work of single users \cite{wellner1991digitaldesk, wellner1993interacting, mackay1999designing} or in collaborative settings \cite{rekimoto1999augmented}. Later the Microsoft Kinect and further commodity depth sensors gave rise to a series of explorations with projector-camera systems.


For example, Xiao et al. \cite{xiao2013worldkit} introduced WorldKit, to allow users to sketch and operate user interface elements on everyday surfaces. Corsten et al. \cite{corsten2013instant} proposed a pipeline for repurposing everyday objects as input devices. Henderson and Feiner also proposed to utilize passive haptic feedback from everyday objects to interact with virtual control elements such as virtual buttons \cite{henderson2010opportunistic}.

Mistry and Maes \cite{mistry2009sixthsense} utilized a necklace-mounted projector-camera system to sense finger interactions and project content on hands or the environment. Following suite, Harrison et al., \cite{harrison2011omnitouch} introduced OmniTouch, a wearable projector-depth-camera system that allowed to project user interface elements on body parts, such as the hand (e.g., a virtual dial pad) or to augment paper using touch. 

\begin{figure}
\includegraphics[width=\columnwidth]{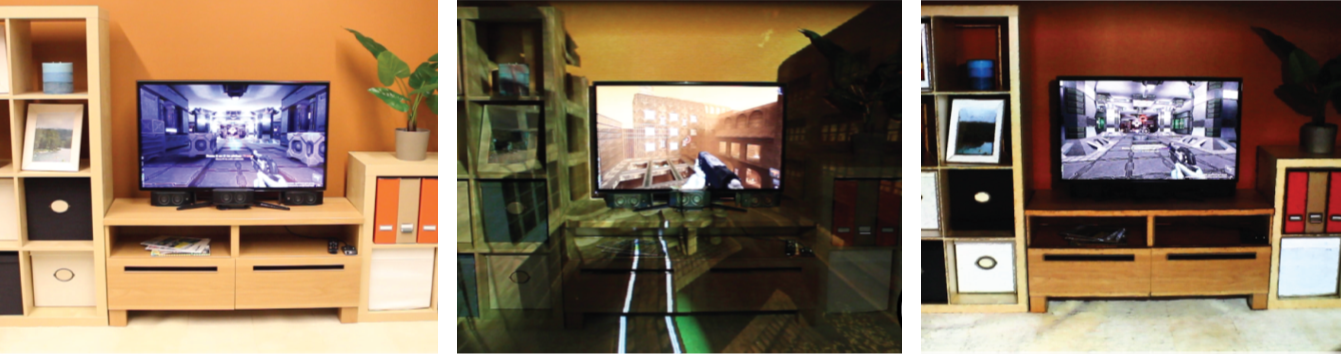}
\caption{The area around a TV is augmented using a projector in IllumiRoom \cite{jones2013illumiroom}. Left: The unmodified scene. Middle: the The field of view of a game is extended beyond the TV boundaries. Right: the appearance of the scene is changed using a cartoon style rendering. Image courtesy of Microsoft Research.}
\label{fig:illumiroom}
\end{figure}

\begin{figure}
\includegraphics[width=0.5\columnwidth]{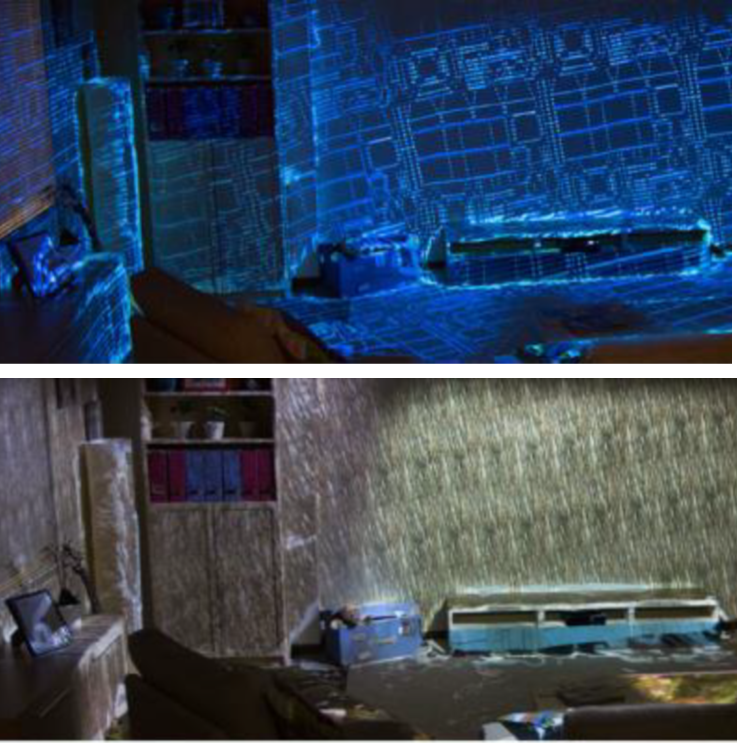}
\caption{Using multiple projector-camera systems a whole room is turned into an immersive experience in RoomAlive \cite{jones2014roomalive}. Two different visualizations of the same physical room can be seen. Image courtesy of Microsoft Research.}
\label{fig:roomalive}
\end{figure}

Further, the idea of interacting with augmented surfaces was later expanded to cover bend surfaces \cite{benko2012miragetable}, walls \cite{jones2013illumiroom}, complete living rooms \cite{jones2014roomalive} or even urban facades \cite{boring2011multi, fischer2012urban}. 
For example, in IllumiRoom \cite{jones2013illumiroom}, the area around a television was augmented using a projector, after initially scanning it with a depth camera, see Figure \ref{fig:illumiroom}. Possible augmentations included extending the field of view of on-screen content, selectively rendering scene elements of a game or changing the appearance of the whole environment using non-photorealistic rendering (e.g., cartoon style or a wobble effect).
In RoomAlive, multiple projector-depth camera units were used to create a 3D scan of a living room as well as to spatially track the user's movement within that room, see Figure \ref{fig:roomalive}. Users were able to interact with digital elements projected in the room using touch and in-air gestures. Apart from entertainment purposes, this idea was also investigated in productivity scenarios such as collaborative content sharing in meetings \cite{fender2017meetalive}. Finally, the augmentation of shape changing interfaces was also explored \cite{rasmussen2012shape, follmer2013inform, leithinger2013sublimate}.
For example, in Sublimate \cite{leithinger2013sublimate} an actuated pin display was combined with a stereoscopic see-through screen to achieve a close coupling between physical and virtual object properties, e.g., for visualizing height fields or NURBS surface modelling. InForm \cite{follmer2013inform} expanded this idea to allow both for user input on its pins (e.g., utilizing them as buttons or handles) as well as manipulation of external objects (such as moving a ball across its surface).

In VR, tangible interaction has been explored using various props. The benefit of using tangibles in VR is that a single physical object can be used to represent multiple virtual objects \cite{araujo2016snake}, even if they show a certain extend of discrepancy. Simeone et al. \cite{simeone2015substitutional}, presented a model of potential substitutions based on physical objects such as mugs, bottles, umbrellas or a torch. Hettiarachchi et al. \cite{hettiarachchi2016annexing} transferred this idea to AR. Harley et al. \cite{harley2017tangible}, proposed a system for authoring narrative experiences in VR using tangible objects.

\section{Gesture-based Interaction}

Touch and in-air gestures and postures make up a large part of interpersonal communication and have also been explored in depth in Mixed Reality. A driver for gesture-based interaction was the desire for "natural" user interaction, i.e. interaction without the need to explicitly handle artificial control devices, but to rely on easy to learn interaction with (to the user) invisible input devices.  While many gesture sets have been explored by researchers or users \cite{piumsomboon2013user}, it can be debated how "natural" those gesture-based interfaces really are \cite{norman2010natural}, e.g., due to the poor affordances. 

Still, the prevalence of small sensors such as RGB and depth cameras, inertial measurement units, radars or magnetic sensors in mobile devices, AR and VR HMDs, as well as continuing advances in hand \cite{barsoum2016articulated,li2019survey}, head \cite{murphy2008head} and body pose estimation \cite{chen2013survey,liu2015survey,gong2016human,sarafianos20163d,dang2019deep,caserman2019survey,chen2020monocular} gave rise to a wide variety of gesture-based interaction techniques being explored for Mixed Reality. 

For mobile devices researchers, began investigating options for interaction next to~\cite{Oakley:2014:IEO:2556288.2557138}, 
above~\cite{kratz2009hoverflow, Freeman:2014:TUA:2628363.2634215}, 
behind~\cite{DeLuca:2013:BAS:2470654.2481330, Wigdor:2007:LTS:1294211.1294259}, 
across~\cite{Schmidt:2012:CIS:2317956.2318005,chen2014duet, radle2014huddlelamp, grubert2017towards}, or 
around~\cite{Zhao:2014:SDI:2642918.2647380,xiao2014toffee} 
the device. 

The additional modalities are either substituting or complementing the devices' capabilities. These approaches typically relied on modifying existing devices using a variety of sensing techniques, which can limit their deployment to mass audiences. Hence, researchers started to investigate the use of unmodified devices. Nandakumar et al. \cite{nandakumar2016fingerio} proposed to use the internal microphones of mobiles to determine the location of finger movements on surfaces, but could not support mid-air interaction. Song et. al \cite{song2014air} enabled in-air gestures using the front and back facing cameras of unmodified mobile devices.  With Surround See, Yang et al.~\cite{yang2013surround} modified the front-facing camera of a mobile phone with an omnidirectional lens, extending its field of view to $360^\circ$ horizontally. They showcased different application areas, including peripheral environment, object and activity detection, including hand gestures and pointing, but did not comment on the recognition accuracy. In GlassHands, it was demonstrated how the input space around a device can be extended, by using a built-in front-facing camera of an unmodified handheld device and some reflective glasses, like sunglasses, ski goggles or visors \cite{grubert2016glasshands}. This work was later extended to investigate the feasibility of utilizing eye reflections \cite{schneider2017poster, schneider2017towards}.

While being explored since the mid 90's in tabletop-based AR \cite{crowley1995finger, brown2000finger, dorfmuller2001finger}, for handheld AR, vision-based finger and hand tracking became popular since the mid 2000's \cite{lee2007handy, lee20083d, shen2011vision, hurst2013gesture}. Yusof et al. \cite{yusof2016review} provide a survey on the various flavors of gesture-based interaction in handheld AR, including marker-based and marker-less tracking of fingers or whole hands.

An early example for in-air interaction using AR HMDs is presented by Kolsch et al. \cite{kolsch2004vision}, who demonstrated finger tracking with a head-mounted camera. Xiao et al. \cite{xiao2018mrtouch} showed how to incorporate touch gestures on everyday surfaces in to the Microsoft HoloLens. Beyond hand and finger tracking, full-body tracking using head-mounted cameras was also explored \cite{cha2018towards}. Also, reconstruction of facial gestures, e.g., for reenactment purposes, when wearing HMDs has seen increased interest \cite{chen2018real,elgharib2019egoface, thies2018facevr, zollhofer2018state}. Further solutions for freehand interaction were also proposed including a wrist-worn gloveless sensor \cite{kim2012digits}, swept frequency capacitive sensing \cite{sato2012touche}, an optical mouse sensor attached to a finger \cite{yang2012magic}, or radar-based sensing \cite{wang2016interacting}. 

Many AR and VR in-air interaction techniques rely on using arms not being supported by a surface (e.g. an elbow resting on a table). Hence, to facilitate reliable selection, targets are designed to be sufficiently large and spaced apart \cite{speicher2018VRselection}. Also, while the addition of hand tracking to modern AR and VR HMDs allows for easy access to in-air gestures, the accuracy of those spatial tracking solutions still is significantly lower than dedicated lab-based external tracking systems \cite{schneider2020accuracy}.

Besides interaction with handheld or head-worn devices, also whole environments such as rooms can be equipped with sensors to facilitate gesture-based interaction \cite{branzel2013gravityspace, nabil2017interactive, zhang2018wall++}. In VR, off-the-shelf controllers were also appropriated to reconstruct human poses in real-time \cite{jiang2016real, caserman2019real}.

\section{Pen-based Interaction}

In-air interactions in AR and VR typically make use of unsupported hands or controllers designed for gaming. In addition, pens (often in combination with tablets as supporting surface) have also been explored as input device. Szalavári and Gervautz \cite{szalavari1997personal}, and, similarly, Billinghurst et al. \cite{billinghurst19973d} utilized pens for input on physical tablets in AR respectively VR. Watsen et al. \cite{watsen1999handheld} used a handheld Personal Digital Assistant (PDA) for operating menus in VR. In the \textit{Studierstube} framework, pens were used to control 2D user interface elements on a PDA in AR.  Poupyrev et al. \cite{poupyrev1998virtual} used a pen for notetaking in VR.  Gesslein et al. \cite{gesslein2020pen} used a pen for supporting spreadsheet interaction in Mobile VR.

\begin{figure}
\includegraphics[width=\columnwidth]{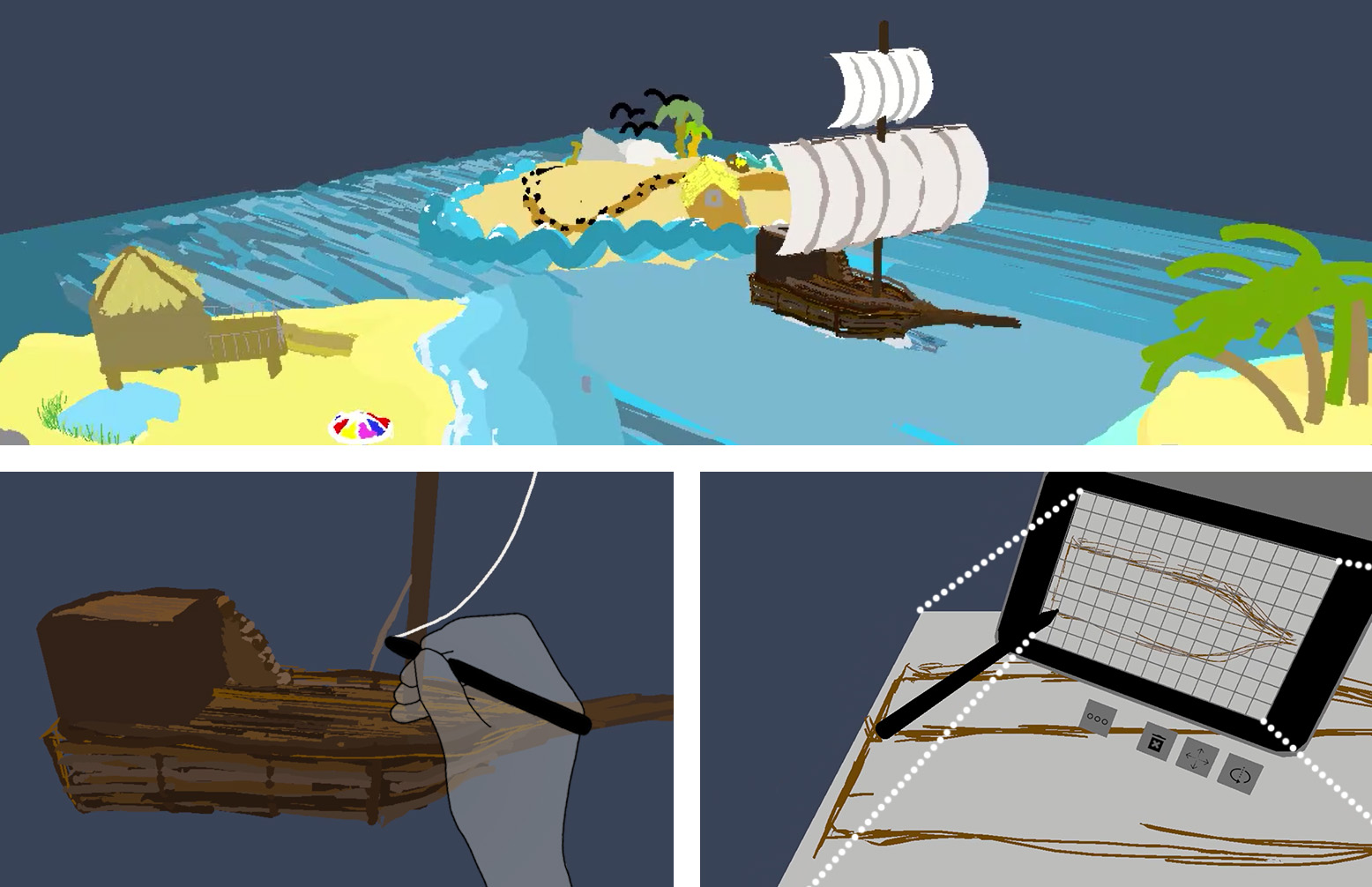}
\caption{Sketching in VR (top) supported by pen-based input in mid-air (bottom left) as well as using a tablet as supporting 2D surface (bottom right) \cite{drey2020sketching}. Image courtesy of Tobias Drey.}
\label{fig:sketchingvr}
\end{figure}

Researches also investigated the use of pens for drawing and modelling, see Figure \ref{fig:sketchingvr}. Sachs et al. \cite{sachs19913} presented an early system of 3D CAD modeling using a pen. Deering \cite{deering1995holosketch} used a pen for in-air sketching in a fishtank VR environment. Keeve et al. \cite{keefe2001cavepainting} utilized a brush for expressive painting in a Cave Automatic Virtual Environment (CAVE) environment. Encarnacao \cite{encarnacao1999translucent} used a pen and pad for sketching in VR on top of an interactive table. Fiorentino et al. \cite{fiorentino2005senstylus} explored the use of pens in mid-air for CAD applications in VR. Xin et al. \cite{xin2008napkin} enabled the creation of 3D sketches using pen and tablet interaction in handheld AR. Yee et al. \cite{yee2009augmented} used a pen-line device along a VST HMD for in-situ sketching in AR. Gasquez et al. \cite{gasques2019pintar, gasques2019you}, Arora et al. \cite{arora2018symbiosissketch}, as well as Drey et al. \cite{drey2020sketching} noted the benefits of supporting both free-form in-air sketching as well as sketching on a supporting 2D surface in AR and VR. Suzuki et al. \cite{suzuki2020realitysketch} expanded previous sketching applications for AR with dynamic and responsive graphics, e.g. to support physical simulations.

The performance of pen-based input was also investigated in VR. Bowman and Wingrave \cite{bowman2001design} compared pen and tablet input for menu selection against floating menus and a pinch-based menu system and found that pen and tablet interaction was significantly faster. Teather and Stuerzlinger \cite{teather2011pointing} compared pen-based in put to mouse input for target selection in a fishtank VR environment and found that 3D pointing was inferior to 2D pointing when targets where rendered stereoscopically. Arora et al. \cite{arora2017experimental} compared pen-based mid-air painting to surface-supported painting and found supporting evidence that accuracy improved using a physical drawing surface. Pham et al. \cite{pham2019pen} indicated that pens significantly outperform controllers for input in AR and VR and is comparable to mouse-based input for target selection. Batmaz et al. explored different pen grip styles for target selection in VR \cite{batmaz2020precision}. 

\section{Gaze-based Interaction}
Besides utilizing touch input, in-air gestures or physical input devices, gaze has also been explored as input modality in Mixed Reality. Duchowski \cite{duchowski2018gaze} presents a review of 30 years of gaze-based interaction, in which gaze-based interaction is categorized within a taxonomy that splits
interaction into four forms, namely diagnostic (off-line measurement), active (selection, look to shoot), passive (foveated rendering or gaze-contingent displays), and expressive (gaze synthesis).

For VR, Mine \cite{mine1995virtual} proposed to use gaze-directed steering and look-at menus, as early as 1995. Tanriverdi and Jacob \cite{tanriverdi2000interacting} highlighted that VR can benefit from gaze tracking. They stated that physical effort can by minimized through gaze and that user’s natural eye movement can be employed to perform interactions in VR (e.g., with distant objects). They also indicated that a proposed heuristic gaze selection technique outperformed virtual hand-based interaction in terms of task-completion time. Cournia et al. \cite{cournia2003gaze} found that dwell-time based selection was slower than manual ray-pointing. Duchowski et al. \cite{duchowski2001binocular} presented  software techniques for binocular eye tracking within VR as well as their application to aircraft inspection training. Specifically, they presented means for integrating eye trackers into a VR framework, novel 3D calibration techniques and techniques for eye-movement analysis in 3D space. In 2020, Burova et al. \cite{burova2020utilizing} also utilized eye-gaze analysis in industrial tasks. They used VR to develop AR solutions for maintenance tasks and collected gaze data to elicit comments from industry experts on the usefulness of the AR simulation. Zeleznik et al. \cite{zeleznik2005look} investigated gaze interaction for 3D pointing, movement, menu selection and navigation (orbiting and flying) in VR. They introduced \textit{Lazy} interactions that minimize hand movements, \textit{Helping Hand} techniques in which gaze augments hand-based techniques as well as \textit{Hands Down} techniques, in which the hand can operate a separate input device. Piumsomboon et al. \cite{piumsomboon2017exploring} presented three novel eye-gaze-based interaction techniques for VR: \textit{Duo-Reticles}, an eye-gaze selection techniques based on eye-gaze and inertial reticles,  \textit{Radial Pursuit}, a smooth pursuit-based technique for cluttered object  and \textit{Nod and Roll}, a head-gesture-based interaction based on the vestibulo-ocular reflex. 

\section{Haptic Interaction}

Auditory and visual channels are widely addressed sensory channels in AR and VR systems. Still, human experiences can be enriched greatly through touch and physical motion. Haptic devices enable the interaction between humans and computers by rendering mechanical signals to stimulate human touch and kinesthetic channels. Research in haptics has a long tradition and incorporates expertise from various fields such as robotics, psychology, biology and computer science. Haptics also play a role in diverse application domains such as gaming \cite{deng2013survey}, industry \cite{xia2013review}, education \cite{minogue2006haptics} or medicine \cite{westebring2008haptics,coles2010role,hamza2019survey}.
Haptic interactions are based on cutaneous/tactile (i.e. skin-related) and kinesthetic/proprioceptive (i.e. related to the body pose) sensations. Various devices have been proposed for both sensory channels, varying in form factor, weight, mobility, comfort as well as the fidelity, duration and  intensity of haptic feedback. For recent surveys, please see \cite{bermejo2017survey,pacchierotti2017wearable}.

\begin{figure}
\includegraphics[width=\columnwidth]{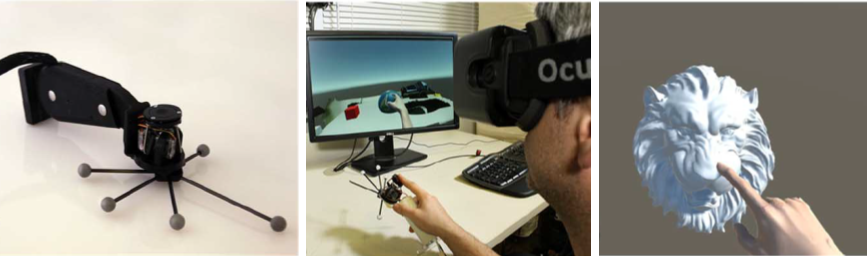}
\caption{A haptic controller for VR (\textit{NormalTouch}, \cite{benko2016normaltouch}) is handheld by a user (middle), allowing to render the surface height and normal when touching a virtual object (right). Image courtesy of Microsoft Research.}
\label{fig:hapticsvr}
\end{figure}

Also, in VR, using haptic feedback has a long tradition \cite{mcneely1993robotic}. A commonly used active haptic device for stationary VR environment with a limited movement range of the users hands, is the PHANToM , which is a grounded system  (or manipulandum) offering a high fidelity but low portability. Hence, over time substantial research efforts have been made in creating mobile haptic devices for VR \cite{pacchierotti2017wearable}, see Figure, \ref{fig:hapticsvr}.

In AR, a challenge for using haptics is that the display typically occludes real objects the user might want to interact with. Also, in OST displays, the haptic device is still visible behind virtual objects rendered on the display. When using VST displays, the haptic device might be removed by inpainting \cite{sandor2007visuo}.

Besides \textit{active} haptic systems, researchers have also investigated the use of low-fidelity physical objects to augment virtual environments in \textit{passive} haptics. An early example of this type of haptic feedback is presented by Insko \cite{insko2001passive}, who showed that passive haptics can improve both sense of presence and spatial knowledge training transfer in a virtual environment. 

\begin{figure}
\includegraphics[width=0.8\columnwidth]{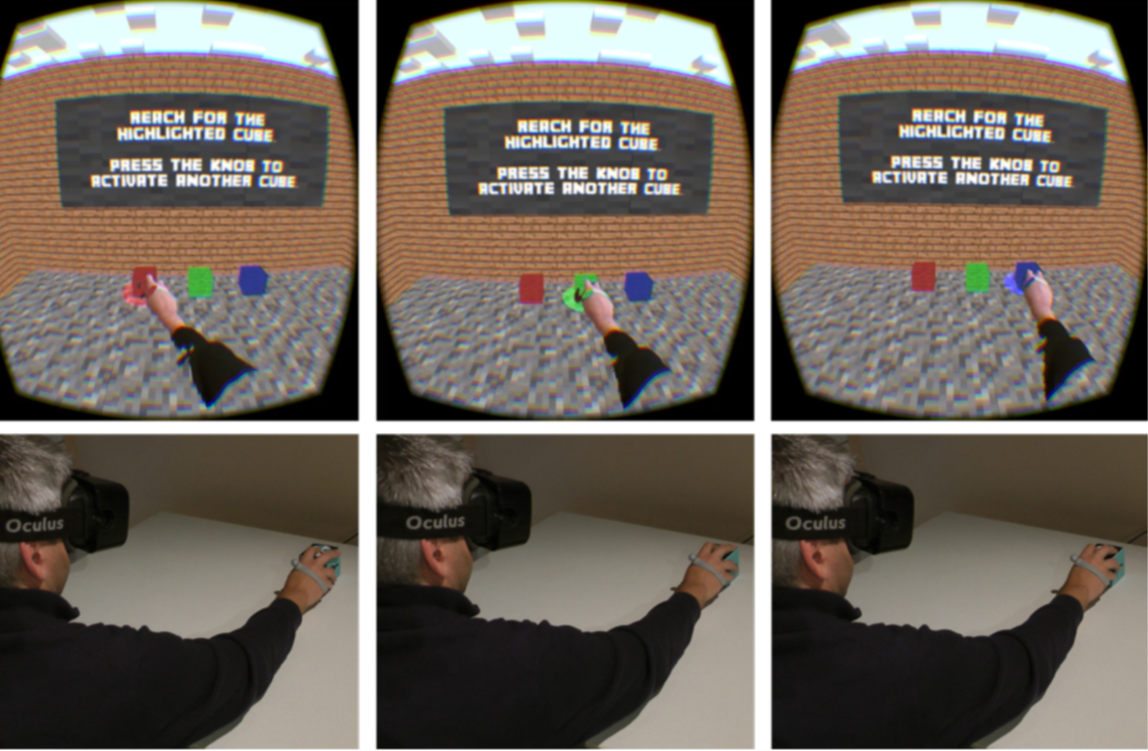}
\caption{A user has the illusion of touching three different virtual cubes (top row), but is solely touching a single physical cube, to which his arm movements are redirected (bottom row) \cite{azmandian2016haptic}. Image courtesy of Microsoft Research.}
\label{fig:hr}
\end{figure}

A challenge when using passive haptic feedback, besides a mismatch in surface fidality, is that the objects used for feedback are typically static. To mitigate this problem two strategies can be employed. First, the objects themselves can be moved during interaction by mounting them on robotic platforms such as robots \cite{suzuki2020roomshift, wang2020movevr} or by human operators \cite{cheng2014haptic, cheng2015turkdeck}. Second, the movements of the user themselves can be redirected to a certain extend by decoupling the physical motion of a user from the perceived visual motion.
This can be done with individual body parts such as hands \cite{azmandian2016haptic,cheng2017sparse}, see Figure \ref{fig:hr} or the whole body using redirected walking techniques \cite{kohli2005combining, nilsson201815}.

\section{Multimodal Interaction}


While, often, AR and VR system offer single input channels along with audio-visual output, rich interaction opportunities arise when considering the combination of further input and output modalities. Complementing the strengths of multiple channels can lead to enriched user experiences. While multimodal (or multisensory) output is typically concerned with increasing the immersion and sense of presence in a scene, multimodal input typically tries to increase the efficiency of user interaction with a AR or VR system. For overviews about mutlimodal interaction beyond AR and VR, please see works by Jaimes and Sebe \cite{jaimes2007multimodal} or Turk \cite{turk2014multimodal}. Nizam et al. also provide a recent overview about multimodal interaction for specifically for AR \cite{nizam2018review}.

The use of multisensory output such as the combination of audiovisual output with smell and touch has been shown to increase presence and perceived realism in VR \cite{cater1994smell, hoffman1998physically} and has been employed as early as in the 1960s \cite{heilig1962sensorama}. Gallace et al. discussed both benefits and challenges when utilizing multiple output modes in VR \cite{gallace2012multisensory}. Extrasensory experiences, \cite{lemmon1937extra, dublon2014extra} (such as making temperature visible through infrared cameras) has also been explored \cite{knierim2018look}. 

In AR, Narumi et al. \cite{narumi2012augmented} showed that increasing the perceived size of a real cookie using AR also increased the feeling of satiety. Narumi et al. \cite{narumi2011augmented} also created a multisensory eating expeirence in AR by changing the apparent look and smell of cookies. Koizumi et al. \cite{koizumi2011chewing} were able to modulate the perceived food texture using a bone-conducting speaker. Ban et al. \cite{ban2013augmented}, showed that it is possible to influence fatigue while handling physical objects by affecting the perceived weight of those objects through modulating their size using AR.

Regarding multimodal input in VR, the combination of speech and gestures is a commonly used input combination. In 1980, Bolt \cite{bolt1980put} introduced \textit{put-that-there}. Users could immerse themselves in a \textit{Media Room} to place objects within that environment through a combination of gestures and speech. In 1989, Hauptmann \cite{hauptmann1989speech} showed that users preferred a combination of speech and and gestures for the spatial manipulation of 3D objects. Cohen et al. \cite{cohen1999multimodal} used a handheld computer along with speech and gesture for supporting map-based tasks on a virtual workbench. LaViola \cite{laviola1999whole}, used hand-based interaction (sensed through a data glove) along with speech for interior design in VR. Ciger et al. \cite{ciger2003magic} combined speech with pointing of a magic wand on an immersive wall to create "magical" experiences. Burdea et al. \cite{burdea1996multimodal}, present an early survey on VR input and output devices as well as an overview about studies that quantify the potentials of several modalities on simulation realism and immersion. Prange et al. \cite{prange2018medical}, studied the use of speech and pen-based interaction in a medical setting.

In AR, Olwal et al. \cite{olwal2003senseshapes}, combined speech and gestures for object selection. Kaiser et al. \cite{kaiser2003mutual} extended that work by introducing mutual disambiguation to improve selection robustness. Similarly, Heidemann et al. \cite{heidemann2004multimodal}, presented an AR system for online acquisition of visual knowledge and retrieval of memorized objects using speech and deictic (pointing) gestures. Kolsch et al. \cite{kolsch2004vision}, combined speech input with gestures in an outdoor AR environment. Piumsomboon \cite{piumsomboon2014grasp}, studied the use of gestures and speech vs gestures only for object manipulation in AR. They found, that the multimodal was not substantially better than gesture-only based interaction for most tasks (but object scaling). This indicates, that multimodality per se is not always beneficial for interaction, but needs to be carefully designed to suit the task at hand. Rosa et al. \cite{rosa2016re}, discussed different notions of AR and Mixed Reality as well as the role of multimodality. Wilson et al. \cite{wilson2012steerable} used a projector-camera system mounted on a pan-tilt platform for multimodal inteaction in a physical room using a combination of speech and gestures.

\begin{figure}
\includegraphics[width=\columnwidth]{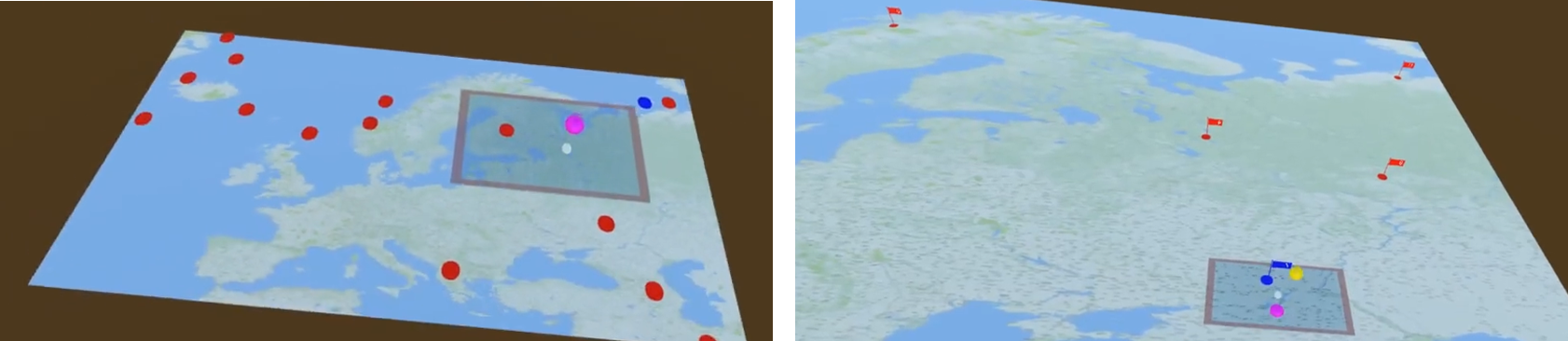}
\caption{Combined in-air and touch interaction for multiscale map navigation \cite{menzner2020above}. Left: View on a virtual map at a small scale. In a study, users were asked to select an active target (blue dot), with inactive targets visualized as red dots. The pink sphere indicates the fingertip position, the white disc below the pivot point for zooming. The touch screen area of the smartphone is shown as semi-transparent blue rectangle with red border. Right: View on the VR scene at 1:1 scale. The circular targets are complemented with 3D flag symbols to indicate that targets can be selected. The two fingertips used for zooming are indicated by yellow and pink spheres. The pivot point indicated by the white disc is visualized half-way between both fingertips.}
\label{fig:multiscale}
\end{figure}

The combination of touch and 3D movements has also been explored in VR and AR. Tsang et al. \cite{tsang2002boom}, introduced the Boom Chameleon, touch display mounted on a tracked mechanical boom and used joint gesture, speech and viewpoint input in a 3D annotation application. Benko et al. \cite{benko2005cross} combined on surface and in-air gestures for content transfer between a 2D screen and 3D space. Mossel et al. \cite{mossel20133dtouch} as well as Marzo et al. \cite{marzo2014combining}, combined touch input and handheld device movement for 3D object manipulations in mobile AR. Polvi et al. \cite{polvi2016slidar} utilized touch and the pose of a handheld touchscreen for remided object positioning in mobile AR. Grandi et al. \cite{grandi2017design}, studied the use of touch and the orientation of a smartphone for collaborative object manipulation in VR. Surale et al. \cite{surale2019tabletinvr} explored the use of touch input on a spatially tracked tablet for object manipulations in VR.  In VR, Menzner et al. \cite{menzner2020above} utilized combined in-air and touch movements on and above smartphones for efficient navigation of multiscale information spaces, see Figure \ref{fig:multiscale}. Several authors combined pen input both in mid-air as well as on touch surfaces to enhance sketching in VR \cite{drey2020sketching} and AR \cite{arora2018symbiosissketch, gasques2019pintar, gasques2019you}.

\begin{figure}
\includegraphics[width=\columnwidth]{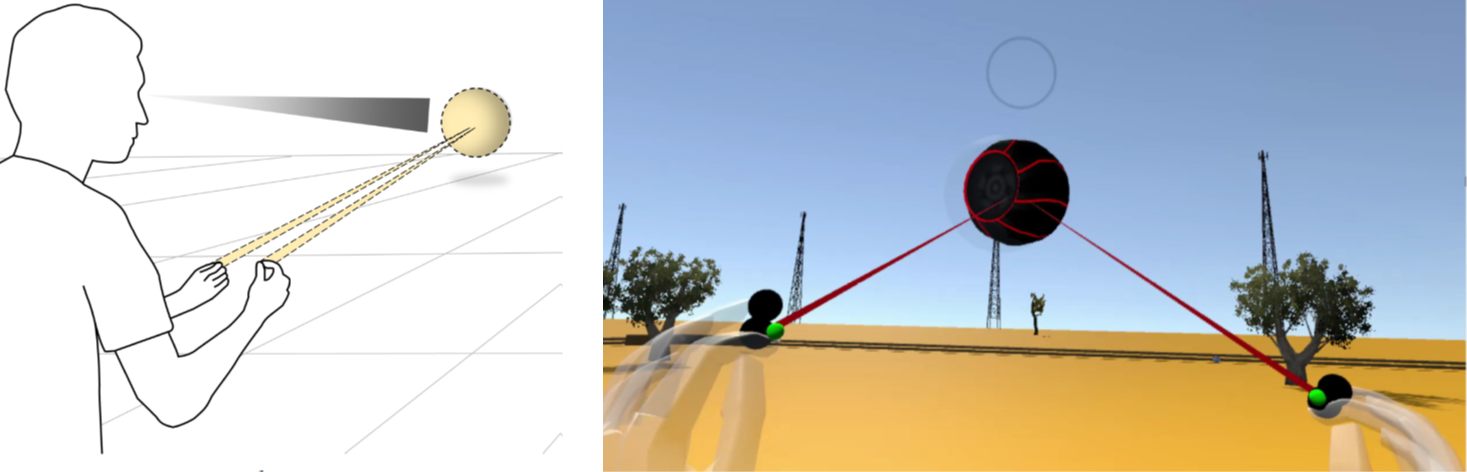}
\caption{Multimodal target selection in VR using a combination of gaze and gestures \cite{pfeuffer2017gaze+}. Image courtesy of Ken Pfeuffer.}
\label{fig:gazepinch}
\end{figure}

\begin{figure}
\includegraphics[width=\columnwidth]{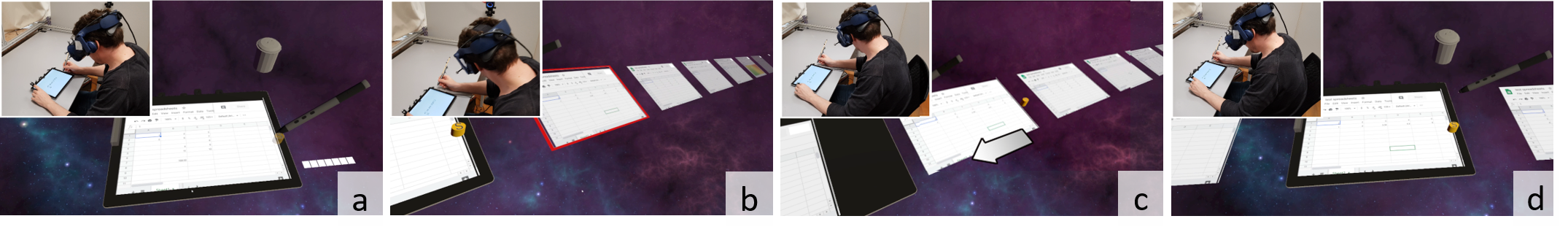}
\caption{Interacting with multiple sheets using a combination of pen-based and gaze-based interaction \cite{gesslein2020pen}. Initially, solely icons indicating the existent of additionally accessible sheets are visible (a). Neighboring sheets are expanded and each sheet the user gazes at is highlighted with a red frame (b). The user taps with his non-dominant hand on the tablet bezel, causing the selected sheet to slide towards the tablet (c), where the user can edit it using the tablet’s touchscreen (d).}
\label{fig:spreadsheet}
\end{figure}

\begin{figure}
\includegraphics[width=\columnwidth]{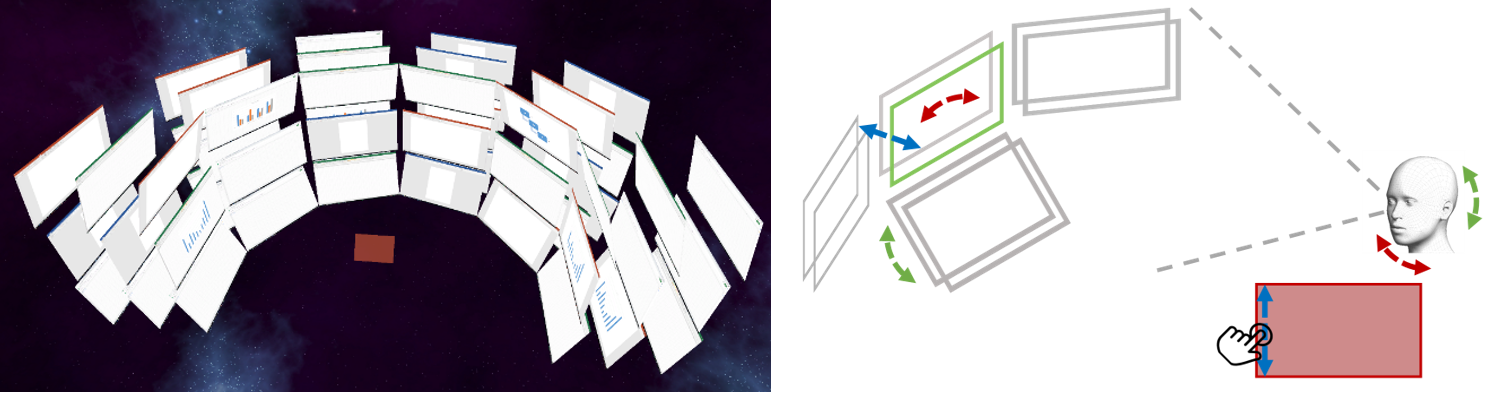}
\caption{Left: a virtual multi-display environment \cite{biener2020breaking}. Right: Users can navigate across displays by gaze (left, right, up, down) and touch (depth).}
\label{fig:bts}
\end{figure}

Also, the combination of eye-gaze with other modalities such as mid-air gestures and head-movements has seen recent interest for interaction in AR and VR. For example, Pfeuffer et al. \cite{pfeuffer2017gaze+} investigated the combination of gaze and gestures in VR, see Figure \ref{fig:gazepinch}. They described \textit{Gaze + Pinch}, which integrates eye gaze to select 3D objects, and indirect freehand gestures to manipulate those objects. They explored this technique for object selection, manipulation, scene navigation, menu interaction, and image zooming. Similarly, Ryu et al. \cite{ryu2019gg} introduced a combined grasp eye-pointing technique for 3D object selection. Kyto et al. \cite{kyto2018pinpointing} combined head and eye gaze for improving target selection in AR. Sidenmark and Gellersen \cite{sidenmark2019eye, sidenmark2020bimodalgaze}, studied  different techniques combining eye and head pointing in VR. Gesslein et al. \cite{gesslein2020pen} combined pen-based input with gaze tracking for efficient interaction across multiple spreadsheets, see Figure \ref{fig:spreadsheet}. Biener et al. \cite{biener2020breaking} utilized gaze and touch interaction for navigating virtual multi-display environments, see Figure  \ref{fig:bts}.

\section{Multi-Display Interaction}
Traditionally, output of interactive systems is often limited to a single display. However, multi-display environments from the desktop to gigapixel displays are also increasingly common for knowledge work and complex tasks such as financial trading or factory management as well as for social applications such as second screen TV experiences \cite{grubert2016challenges}. Surveys about multi-display systems and distributed user interfaces have been presented by Elmqvist \cite{elmqvist2011distributed},  Grubert et al. \cite{grubert2016challenges, grubert2015design,  quigley2015perceptual} and Brudy et al. \cite{brudy2019cross}.

Augmented Reality has the potential to enhance interaction with both small and large displays by adding an unlimited virtual screen space or other complementing characteristics like mobility. However, this typically comes at the cost of a lower display fidelity compared to a physical panel display (such as lower resolution, lower contrast, or a smaller physical field of view in OST HMDs). 

\begin{figure}
\includegraphics[width=0.6\columnwidth]{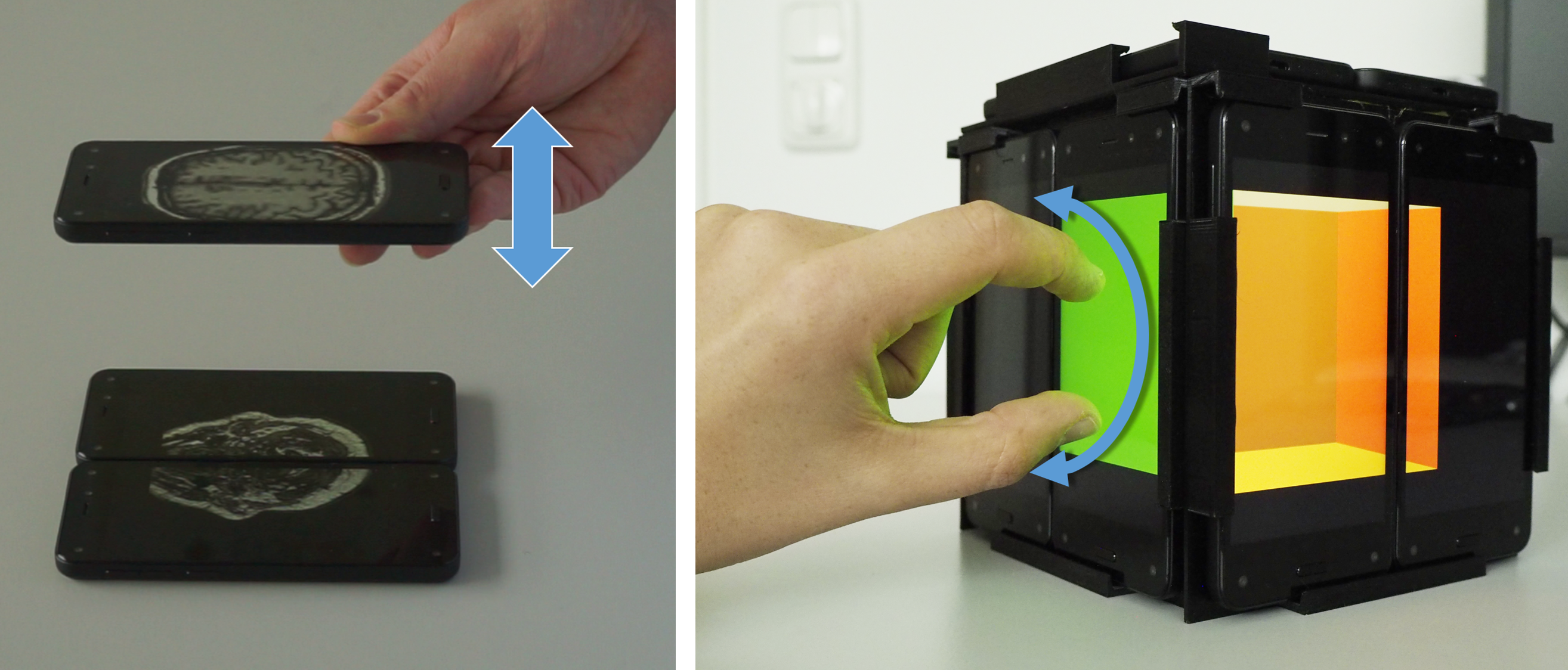}
\caption{Mobile displays can be spatially registered using head-tracking for augmented planar surfaces (left) or creating a fishtank VR display (right) \cite{grubert2017headphones}.}
\label{fig:headphones}
\end{figure}

\begin{figure}
\includegraphics[width=0.5\columnwidth]{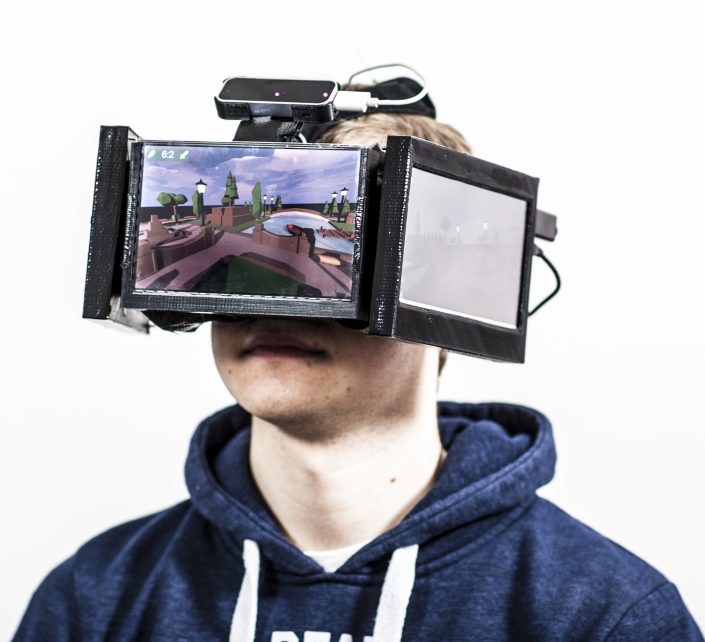}
\caption{Multiple Displays around a users head for enhanced interaction in VR \cite{gugenheimer2017facedisplay}. Image courtesy of Jan Gugenheimer.}
\label{fig:facedisplay}
\end{figure}

\begin{figure}
\includegraphics[width=0.5\columnwidth]{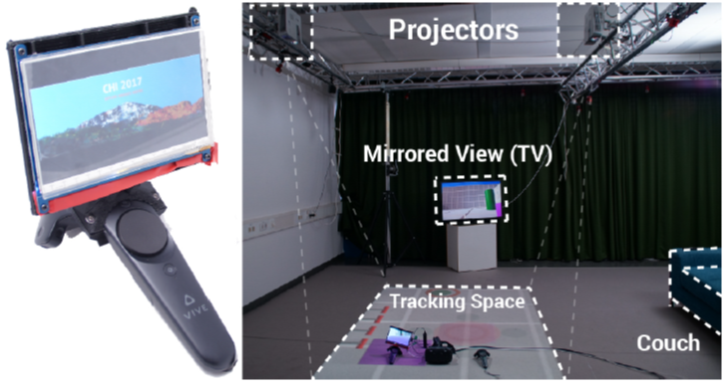}
\caption{Left: A controller mounted dipslay to allow users outside of VR to interact with a user wearing an immersive head-mounted display in ShareVR \cite{gugenheimer2017sharevr}. Right: The ShareVR environment resembling a living room. Image courtesy of Jan Gugenheimer.}
\label{fig:sharevr}
\end{figure}

In 1991, Feiner et al. \cite{feiner1991hybrid}, proposed a hybrid display combining a traditional desktop monitor with an OST HMD and explored a window manager application. Butz et al. \cite{butz1999enveloping}, combined multiple physical displays ranging from handheld to wall-sized with OST HMDs in a multi-user collaborative environment. Baudisch et al. \cite{baudisch2001focus} used a projector to facilitate focus and context interaction on a desktop computer. MacWilliams et al. \cite{macwilliams2003herding} proposed a multi-user game in which players could interact with a tabletop, laptop as well as handheld displays. Serrano et al. \cite{serrano2015gluey} propose to use an OST HMD to facilitate content transfer between multiple physical displays on a desktop. Boring et al. \cite{boring2010touch} used a smartphone to facilitate content transfer between multiple stationary displays. They later extended the work to manipulate screen content on stationary displays \cite{baur2012virtual} and interactive facades \cite{boring2012making} using smartphones. Raedle et al. \cite{radle2014huddlelamp} supported interaction across multiple mobile displays through a top-mounted depth-camera. Grubert et al. \cite{grubert2017headphones, grubert2017towards} used face tracking to allow user interaction across multiple mobile devices, which could be dynamically re-positioned, see Figure \ref{fig:headphones}, left. They also proposed to utilize face tracking \cite{grubert2017headphones, grubert2017mpcubee} for creating a cubic VR display with user perspective rendering, see Figure \ref{fig:headphones}, right. Butscher et al. \cite{butscher2018clusters} explored the combination of VST HMDs with a tabletop displays for information visualization. Reipschl{\"a}ger et al. \cite{reipschlager2019designar, reipschlager2020augmented} combined a high resolution horizontal desktop display with an OST HMD for design activities. Gugenheimer et al.~\cite{gugenheimer2016facetouch} introduced face touch, which allows interacting with display-fixed user interfaces (using direct touch) and world-fixed content (using raycasting). This work was later extended to utilize three touch displays around the user's head \cite{gugenheimer2017facedisplay}, see Figure \ref{fig:facedisplay}. Gugenheimer et al. also introduced \textit{ShareVR} \cite{gugenheimer2017sharevr}, which enabled multi-user and multi-display interaction across users inside and outside of VR, see Figure\ref{fig:sharevr}.  

\begin{figure}
\includegraphics[width=0.5\columnwidth]{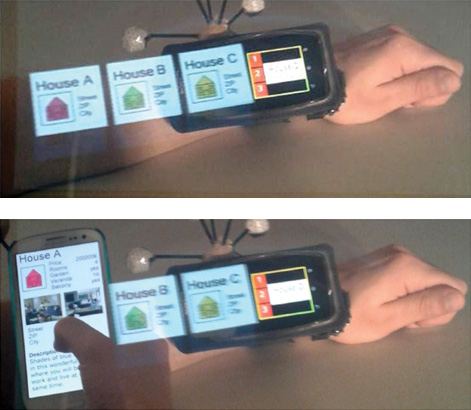}
\caption{Top: the field of view of a smartphone is extended by an optical see-through display, in MultiFi \cite{grubert2015multifi}. Bottom: A smartwatch is used to select individual icons using touch.}
\label{fig:multifi}
\end{figure}

A number of systems also concentrated on the combination of HMDs and handheld as well body-worn displays such as smartwatches, smartphones and tablets in mobile contexts. Here, typically the head-mounted display extends the field of view of the handheld display to provide a larger virtual field of view. In MultiFi \cite{grubert2015multifi}, an OST HMD provides contextual information for higher resolution touch-enabled displays (smartwatch and smartphone), see Figure \ref{fig:multifi}. The authors explored different spatial reference systems such as body-aligned, device-aligned, and side-by-side modes. Similar explorations have followed suit using video-see-through HMDs \cite{normand2018enlarging}, an extended set of interaction techniques \cite{zhu2020bishare}, using smartwatches \cite{wenig2017watchthru, luwatchar, wolf2018performance}, or with a focus on understanding smartphone-driven window management techniques for HMDs \cite{ren2020understanding}. 

Purely virtual multi-display environments have also been explored in AR and VR. In 1993, Feiner et al. \cite{feiner1993windows} introduced head-surrounding and world reference frames for positioning 3D windows in VR. In 1998, Billinghurst et al. \cite{billinghurst1998wearable} introduced the spatial display metaphor, in which information windows are arranged on a virtual cylinder around the user. Since then, virtual information displays have been explored in various reference systems, such as world-, object-, head-, body- or device-referenced systems~\cite{laviola20173d}.  Specifically, interacting with windows in body-centered reference systems~\cite{wagner2013body} has attracted attention, for instance to allow fast access to virtual items \cite{li2009virtual, chen2012extending}, mobile multi-tasking \cite{ens2014personal, ens2014ethereal} and visual analytics~\cite{ens2016spatial}. Lee et al.~\cite{lee2018projective} investigated positioning a window in 3D space using a continuous hand gesture. Petford et al.~\cite{petford2018pointing} compared the selection performance of mouse and raycast pointing in full coverage displays (not in VR). Jetter et al.~\cite{jetter2020vr} proposed to interactively design a space with various display form factors in VR.

\section{Interaction using Keyboard and Mouse}

Being the de facto standard for human-computer interaction in personal computing environments for decades, standard input peripherals such as keyboard and mouse, while initially used in projection-based CAVE environments, were soon replaced by special purpose input devices and associated interaction techniques for AR and VR (see previous sections) . This was partly due to the constraints of those input devices making them challenging to use for spatial input with six degrees of freedom. Physical keyboards typically support solely symbolic input. Standard computer mice are restricted to two-dimensional pointing (along with button clicks, and a scroll-wheel). However, with modern knowledge workers still relying on the efficiency of those physical input devices, researchers revisited how to use them within AR and VR.

With increasing interest in supporting knowledge work using AR and VR HMDs \cite{grubert2018office, ruvimova2020transport, guo2019mixed, li2019holodoc}, keyboard and mouse interaction drew attention by several researchers. 

The keyboard was designed for rapid entrance of symbolic information, and although it may not be the best mechanism developed for the task, its familiarity that enabled good performance by users without considerable learning efforts kept it almost unchanged for many years. However, when interacting with spatial data, they are perceived as falling short of providing efficient input capabilities \cite{besanccon2017mouse}, even though they are successfully used in many 3D environments (such as CAD or gaming \cite{stuerzlinger2011value}), can be modified to to allow 3D interaction \cite{ware1997selection, perelman2015roly} or can outperform 3D input devices in specific tasks such as 3D object placement \cite{berard2009did, sun2018comparing}. Also for 3D object manipulation in AR and VR they found to be not significantly slower than a dedicated 3D input device \cite{krichenbauer2017augmented}.

\begin{figure}
\includegraphics[width=\columnwidth]{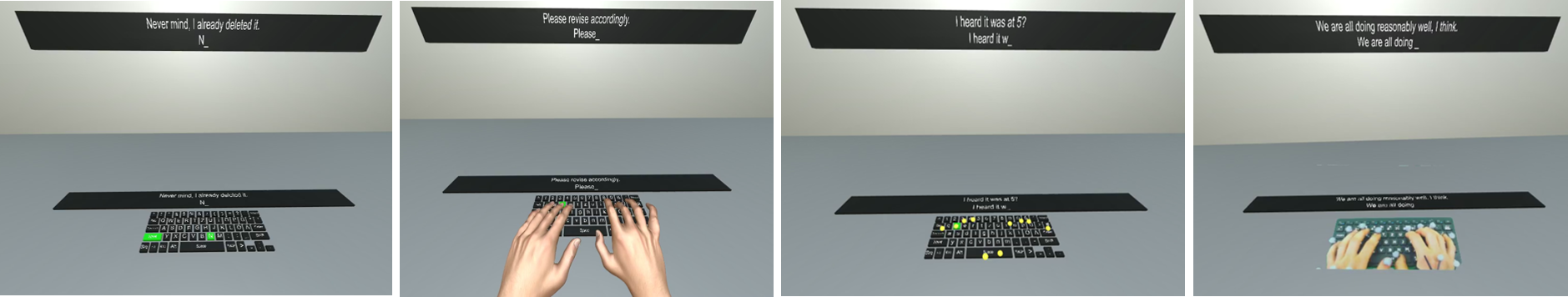}
\caption{Text entry in VR using standard physical keyboards using different hand reprensentations \cite{grubert2018effects}. From left to right: no representation, inverse-kinematic model,  finger tip representation using spheres, video pass-through of the user's hands.}
\label{fig:textentryvr}
\end{figure}

In VR, a number of works investigated the costs of using physical keyboards for standard text entry tasks, see Figure \ref{fig:textentryvr}. Grubert et al. \cite{grubert2018text, grubert2018effects}, Knierim et al. \cite{knierim2018physical} as well as McGill et al. \cite{mcgill2015dose} found physical keyboards to be mostly usable for text entry in immersive head-mounted display-based VR but varied in their observations about the performance loss when transferring text entry from the physical to the virtual world. Pham et al. \cite{pham2019hawkey} deployed a physical keyboard on a tray to facilitate mobile text entry. Apart from standard QWERTY keyboards a variety of further text entry input devices and techniques have been proposed for VR, see \cite{dube2019text}.

Besides using unmodified physical keyboards, there have been several approaches in extending the basic input capabilities of physical keyboard beyond individual button presses. Specifically, input on, above and around the keyboard surface have been proposed using acoustic \cite{kurosawa2013keyboard, kato2010surfboard}, pressure \cite{dietz2009practical, zagler2003fasty, loy2005development}, proximity  \cite{taylor2014type}, capacitive sensors \cite{fallot2006touch, habib2009dgts, tung2015flickboard, rekimoto2003presense, shi2018gestakey, block2010touch}, cameras \cite{wilson2006robust, kim2014retrodepth, ramos2016keyboard}, body-worn orientation sensors \cite{buschek2018extending} or even unmodified physical keyboards \cite{lee2013multidirectional, zhang2014gestkeyboard}. Besides sensing, actuation of keys has also been explored \cite{bailly2013metamorphe}. Embedding capacitive sensing into keyboards has been studied by various researchers. It lends itself  to detect finger events on and slightly above keys and can be integrated into mass-manufacturing processes. Rekimoto et al. \cite{rekimoto2003presense} investigated capacitive sensing on a keypad, but not a full keyboard. Habib et al. \cite{habib2009dgts} and Tung et al. \cite{tung2015flickboard} proposed to use capacitive sensing embedded into a full physical keyboard to allow touchpad operation on the keyboard surface. Tung et al. \cite{tung2015flickboard} developed a classifier to automatically distinguish between text entry and touchpad mode on the keyboard. Shi et al. developed microgestures on capacitive sensing keys \cite{shi2017gestakey, shi2018gestakey}. Similarly, Zheng et al. \cite{zheng2016finger, zheng2018fingerarc} explored various interaction mappings for finger and hand postures. Sekimoro et al. focused on exploring gestural interactions on the space bar \cite{sekimori2018ex}. Extending the idea of LCD-programmable keyboards \cite{LCDKeysHistory}, Block et al. extended the output capabilities of touch-sensitive, capacitive sensing keyboard by using a top-mounted projector \cite{block2010touch}. Several commercial products have also augmented physical keyboards with additional, partly interactive, displays (e.g., Apple Touch Bar \cite{ATB}, Logitech G19 \cite{LTG19}, Razer Death-Stalker Ultimate \cite{RDS}). 

Maiti et al. \cite{maiti2017preventing} explored the use of randomized keyboard layouts on physical keyboards using an OST display. Wang et al. \cite{wang2020towards} explored the use of an Augmented Reality extension to a desktop-based analytics environment. Specifically, they added a stereoscopic data view using a HoloLens to a traditional 2D desktop environment and interacted with keyboard and mouse across both the HoloLens and the desktop.

\begin{figure}
\includegraphics[width=\columnwidth]{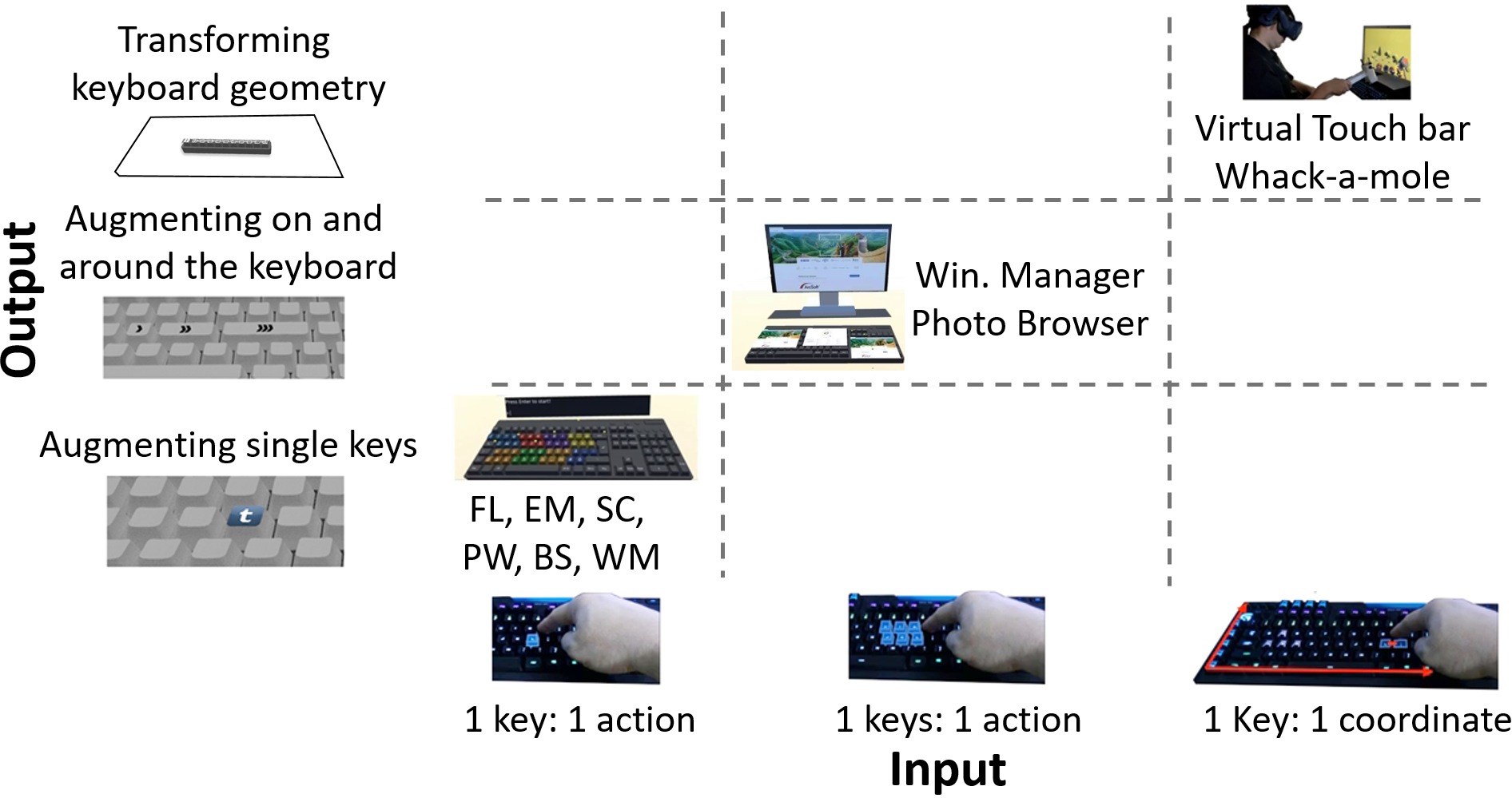}
\caption{Input-output dimensions of reconfiguring physical keyboards in VR with mapped example applications, \cite{schneider2019reconviguration}. The $x$-axis shows input mappings and the $y$-axis shows output mappings. FL: Foreign Languages; EM: Emojis; SC: Special Characters; PW: Secure Password Entry; BS: Browser Shortcuts; WM = Word Macros.
}
\label{fig:reconviguraitonds}
\end{figure}

Schneider et al. \cite{schneider2019reconviguration} explored the design space of using physical keyboards in VR beyond text entry, see Figure \ref{fig:reconviguraitonds}. Specifically, they proposed three different input mappings: 1 key to 1 action (standard mode of interaction using keyboards), multiple keys to a single action (e.g., mapping a large virtual button to several physical buttons), as well as mapping a physical key to a coordinate in a two-dimensional input space. Similarly, they proposed three different output mappings: augmenting individual keys (e.g., showing an emoji on a key), augmenting on and around the keyboard (e.g., adding additional user interface elements on top of the keyboard such as virtual sliders) as well as transforming the keyboard geometry itself (e.g., only displaying single buttons, or replacing the keyboard by other visuals). Those ideas were later also considered in the domain of immersive analytics \cite{grubert2020back}.

Mouse-based pointing has been studied in depth outside of AR and VR for pointing on single monitors \cite{casiez2008impact} as well as multi-display environments \cite{baudisch2004mouse, ashdown2005combining, benko2005multi}. However, it has been found that standard 2D mice do not adapt well to multi-display interaction \cite{waldner2010bridging}, an issue which is also relevant for AR and VR. Consequently, standard mice have been modified in various ways to add additional degrees of freedom. For example, Villar et al. \cite{villar2009mouse} explored multiple form factors for multi-touch enabled mice. Other researchers have added additional mouse sensors to support yawing \cite{mackenzie1997two, olwal2004unit}, pressure sensors for discrete selection \cite{cechanowicz2007augmenting, kim2008inflatable} to allow for three instead of two degrees of freedom. Three-dimensional interaction was enabled using the \textit{Rockin'Mouse} \cite{balakrishnan1997rockin} and the \textit{VideoMouse} \cite{hinckley1999videomouse}. Both works added a dome below the device to facilitate 3D interaction. Steed and Slater \cite{steed19953d} proposed to add a dome on top of the mouse rather than below. Further form factors have also been proposed to facilitate pointing based interaction in 3D \cite{frohlich2000cubic, froehlich2006globefish}. Recently, researchers also worked on unifying efficient input both in 2D and 3D \cite{perelman2015roly, saidi2017tdome}. 

Standard mice using a scroll wheel can also be efficiently used for 3D object selection when being combined with gaze-tracking in virtual multi-display environments \cite{biener2020breaking}. For example, in the Windows Mixed Reality Toolkit \cite{mswmt2019}, the x and y- movement of the mouse can be mapped to the x and y movements on a proxy shape such as a cylinder (or any object on that cylinder, like a window). The scroll wheel is used for changing the pointer depth (in discrete steps). The x- and y- movements can be limited to the current field of view of the user to allow for acceptable control to display ratios. The user gaze can then be used to change the view on different regions of the proxy shape.

\section{Virtual Agents}

Virtual agents can be considered as "intelligent" software programs performing tasks on behalf of users' questions or commands. While it can be argued, what "intelligent" really means in this context, a widely accepted characteristic of this "intelligence" is context-aware behaviour \cite{volkel2020intelligent, grubert2016towards}. This allows an agent to interact with the user and environment through sensing and acting in an independent and dynamic way. The behaviour is typically well-defined and allows to trigger actions based on a set of conditions \cite{nwana1996software}. The rise of voice assistants (or conversational agents) \cite{hoy2018alexa}, which interact with users through natural language, has brought media attention and a prevalence in various areas such as home automation, in-car operation,  automation of call centers, education and training \cite{norouzi2018systematic}. 

\begin{figure}
\includegraphics[width=0.5\columnwidth]{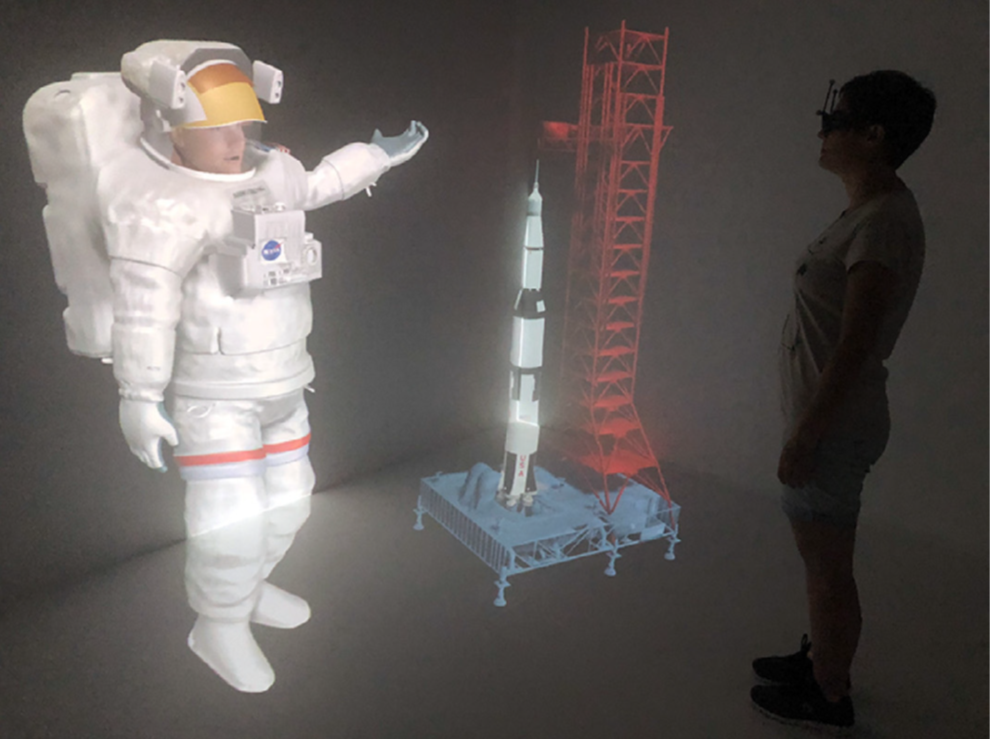}
\caption{An intelligent virtual agent in Virtual Reality \cite{schmidt2019effects}. Image courtesy of Gerd Bruder.}
\label{fig:agentvr}
\end{figure}

\begin{figure}
\includegraphics[width=0.5\columnwidth]{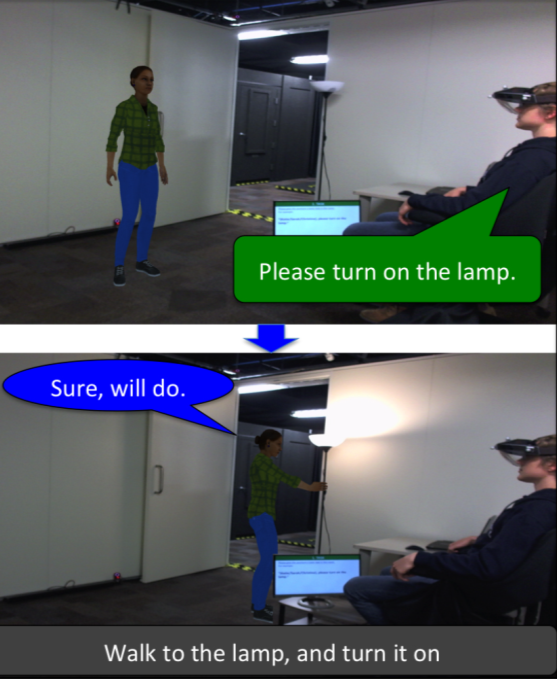}
\caption{An intelligent virtual agent in Augmented Reality \cite{kim2018does}. Top: The user requests to turn of the light. Bottom: the agent walks towards the physical light and turns the wirelessly connected light bulb off.  Image courtesy of Gerd Bruder.}
\label{fig:agentar}
\end{figure}

In AR and VR, virtual agents often use more than a single modality for input and output. Complementary to voice in- an output, virtual agents in AR and VR can typically react to body gestures or postures or even facial expressions of the users. Due to their graphical representations those agents are embodied in the virtual world. 
The level of embodiment of a virtual agent has been studied extensively \cite{dehn2000impact, yee2007meta}. For example it has been shown, that the effect of adding a face was larger than effect of visual realism (both photo-realism and behavioural realism of the avatar) \cite{yee2007meta}. In VR, the level of visual realism of the virtual agent is typically matched to the visual realism of the environment, see Figure \ref{fig:agentvr}. In contrast, in AR, there is often a noticeable difference between the agent representation and the physical scene and those effects are still underexplored \cite{kim2018does}, see Figure \ref{fig:agentar}. Hantono et al. review the use of virtual agents in AR in educational settings. Norouzi et al. provide review of the convergence between AR and virtual agents \cite{norouzi2019systematic}.

Specifically for AR, Maes et al. \cite{maes1997alive} introduced a magic mirror AR system, in which humans could interact with a dog through both voice and gestures. Similarly, Cavazza et al. \cite{cavazza2003interacting} allowed participants to interact with virtual agents in an interactive storytelling environment. MacIntyre et al. \cite{macintyre2001augmented} used pre-recorded videos of phyiscal actors to let users interact with them using OST HMDs. Anabuki et al. \cite{anabuki2000welbo} highlight that having virtual agents and users share the same physical environment is the most distinguishing aspect of virtual agents in AR.  They introduced Welbo, an animated virtual agent, which is is aware of its physical environment and can avoid standing in the user's way. Barakony et al. \cite{barakonyi2004agents} presented \textit{AR Puppet}, a system that explored the context-aware animated agents within AR. The authors investigated aspects like visualization, appearance or behaviors. They also studied AR-specific aspects such as the ability of the agent to avoid physical obstacles or its ability to interact with physical objects. Based on this initial research, the authors explored various applications \cite{barakonyi2005augmented, barakonyi2006ubiquitous}. Chekhlov et al. \cite{chekhlov2007ninja} presented a system based on Simultanteous Localization and Mapping (SLAM) \cite{durrant2006simultaneous}, in which a virtual agent had to move in a physical environment. Blum et al. \cite{blum2012final} introduced an outdoor AR game which included virtual agents. Kotranza et al. \cite{kotranza2008virtual,kotranza2009mixed} used a tangible physical representation of a human that could be touched, along with a virtual visual representation in a medical education context. They called this dual representaiton \textit{mixed reality humans} and argued that affording touch between a human and a virtual agent enables interpersonal scenarios.

\section{Summary and Outlook}
This chapter served as an overview of a wide variety of interaction techniques MR - covering both device- and prop-based input such as tangible interaction, pen and keyboard input as well as utilizing human effector-based input such as spatial gestures, gaze or speech.

The historical development of the presented techniques was closely coupled to the available sensing capabilities. For example, in order to recognize props such as paddles \cite{kato2000virtual}, they had to be large enough in order to let fiducials be recognized by low-resolution cameras. With the advancement of computer vision-based sensing, fiducials could be come smaller, change their appearance to natural looking images or be omitted altogether (e.g., for hand and finger tracking). Further, the combination of more than one modality became possible through increasing computational capabilities of MR systems. 

In the future, we expect an ongoing trend of both minimizing the size and price of sensors, as well as the ubiquitous availability of those sensors, in dedicated computing devices, in everyday objects \cite{weiser1999computer}, on \cite{schmidt2015biosignals} or even in the human body itself \cite{schraefel2019in5}. Hence, MR interaction techniques will play a central role on shaping the future of both pervasive computing \cite{grubert2016towards} as well as augmenting humans with (potentially) super human capabilities (e.g., motor capabilities \cite{kazerooni2008review, kunze2017superhuman}, cognitive and perceptual capabilities \cite{schmidt2017augmenting}). Besides technological and interaction challenges along the way, the field of MR interaction will greatly benefit from including both social and ethical implications when designing future interfaces.

\bibliographystyle{ACM-Reference-Format}
\bibliography{mrinteraction}










\end{document}